\documentclass[conference]{IEEEtran}
\IEEEoverridecommandlockouts
\usepackage{cite}
\usepackage{amsmath,amssymb,amsfonts}
\usepackage{algorithmic}
\usepackage{graphicx}
\usepackage{textcomp}
\usepackage{xcolor}
\usepackage{subcaption} 
\usepackage{booktabs}
\usepackage{todonotes}

\def\BibTeX{{\rm B\kern-.05em{\sc i\kern-.025em b}\kern-.08em
    T\kern-.1667em\lower.7ex\hbox{E}\kern-.125emX}}
\begin{document}

\title{Distributed Resource Selection for Self-Organising Cloud-Edge Systems}

 \author{\IEEEauthorblockN{Quentin Renau}
\IEEEauthorblockA{\textit{School of Computing, Engineering,}\\
\textit{and the Built Environment} \\
\textit{Edinburgh Napier University}\\
Edinburgh, UK \\
q.renau@napier.ac.uk}
\and
\IEEEauthorblockN{Amjad Ullah}
\IEEEauthorblockA{\textit{School of Computing, Engineering,}\\
\textit{and the Built Environment} \\
\textit{Edinburgh Napier University}\\
Edinburgh, UK \\
a.ullah@napier.ac.uk}
\and
\IEEEauthorblockN{Emma Hart}
\IEEEauthorblockA{\textit{School of Computing, Engineering,}\\
\textit{and the Built Environment} \\
\textit{Edinburgh Napier University}\\
Edinburgh, UK \\
e.hart@napier.ac.uk}
}

\maketitle

\begin{abstract}
This paper presents a distributed resource selection mechanism for diverse cloud-edge environments, enabling dynamic and context-aware allocation of resources to meet the demands of complex distributed applications. By distributing the decision-making process, our approach ensures efficiency, scalability, and resilience in highly dynamic cloud-edge environments where centralised coordination becomes a bottleneck. The proposed mechanism aims to function as a core component of a broader, distributed, and self-organising orchestration system that facilitates the intelligent placement and adaptation of applications in real-time. This work leverages a consensus-based mechanism utilising local knowledge and inter-agent collaboration to achieve efficient results without relying on a central controller, thus paving the way for distributed orchestration. Our results indicate that computation time is the key factor influencing allocation decisions. Our approach consistently delivers rapid allocations without compromising optimality or incurring additional cost, achieving timely results at scale where exhaustive search is infeasible and centralised heuristics run up to 30 times slower.
\end{abstract}

\section{Introduction} \label{sec:intro}
The Cloud-Edge continuum constitutes a dynamic and heterogeneous landscape of computational resources distributed across diverse administrative domains, from centralised cloud data centres to distributed edge nodes, each exhibiting varying capabilities, constraints, and performance profiles~\cite{moreschini2022cloud}. The proliferation of such environments has significantly reshaped how modern distributed applications select and use resources to meet performance, latency, and locality requirements. Orchestration systems are often tasked with handling the complexities of deploying and managing these applications, where resource selection is the critical component, responsible for identifying the best-suited resources across the diverse continuum~\cite{ullah2021micado}.

The landscape of resource selection---also referred to as resource or task allocation---is highly diverse, encompassing techniques from P2P and gossip-based~\cite{jimenez2020hydra,sim2022cooperative} to probabilistic and learning-driven approaches~\cite{schneider2021distributed,li2025adaptive,ullah2023optimizing}. These solutions vary in complexity, applicability, and effectiveness based on their design objectives and scope. Nonetheless, when viewed purely from an architectural standpoint, all approaches can be broadly categorised as either centralised or distributed.

Centralised approaches rely on a single orchestrator with a global view of the infrastructure. While effective under stable conditions, it becomes impractical in cloud-edge environments, where infrastructure is highly dynamic, geographically distributed, and spans multiple providers---each with a dynamic view of their infrastructure---thus limiting the feasibility of a global view at the centralised level. Moreover, they suffer from scalability bottlenecks, latency overhead, and a single point of failure, making them unsuitable for time-sensitive or large-scale deployments. 

Distributed approaches, on the other hand, delegate decision-making to multiple autonomous entities that operate based on local information and coordination, offering improved scalability, fault tolerance, and responsiveness. However, designing such mechanisms is challenging due to the need for effective coordination and consistent decision-making across distributed nodes. In this regard, this paper introduces a distributed resource selection mechanism for the cloud–edge continuum, forming a core component of a broader orchestration framework~\cite{UllahMAKKDMWK25, kiss2024swarmchestrate}, responsible for automated application deployment and runtime management. The proposed mechanism empowers autonomous resource agents, each with partial local knowledge, to collaboratively select resources via lightweight coordination and consensus. More specifically, the contributions of this paper are as follows:
\begin{enumerate}
    \item A distributed resource selection mechanism using a state-of-the-art multi-agent task allocation algorithm.
    \item We demonstrate that our approach matches the performance of an optimal centralised exhaustive search while computing allocations significantly faster than two baseline centralised methods.
    \item We show that it effectively scales to large application and resource sets, delivering timely allocations even when exhaustive search is infeasible and centralised heuristics are up to $30$ times slower.
\end{enumerate}
The remainder of this paper is structured as follows. Section~\ref{sec:relatedWork} outlines the broader orchestration framework to which this work contributes, and discusses related research. Section~\ref{sec:methods} details the three core technical methods underpinning our system. Section~\ref{sec:results} presents a comprehensive evaluation, focusing on the efficiency, performance, and scalability of the proposed approach. Finally, Section~\ref{sec:conclusion} concludes this paper. For \textbf{\emph{reproducibility}}, all code, data, and additional plots are publicly available via Zenodo~\cite{dataCBBA}.
\section{Background, Motivation and Related work} \label{sec:relatedWork}
This section provides the necessary background to contextualise our work. We begin by outlining the distributed orchestration framework---Swarmchestrate~\cite{kiss2024swarmchestrate}---within which our proposed mechanism operates, followed by an overview of the adopted implementation technique---distributed task allocation. Finally, we review related work.

\subsection{Swarmchestrate overview} \label{sec:swarmchestrate}
Swarmchestrate, an ongoing Horizon Europe project, is a distributed orchestration framework for the cloud-edge continuum. Its vision is to enable self-organising capabilities that support dynamic, context-aware deployment and management of complex applications across heterogeneous infrastructure, without a central orchestrator. More specifically, it targets three key challenges: first, enabling seamless, concurrent access to a highly heterogeneous resource landscape while coordinating end-to-end services across multiple cloud, fog, and edge providers; second, optimising resource selection to balance key objectives such as performance, cost, and energy efficiency; and third, ensuring trust, reliability, and security in deployments spanning diverse administrative domains, geographic locations, and network environments. This paper contributes to addressing the first two challenges.

The interface of the Swarmchestrate is a P2P network of \textbf{\textit{Resource Agents (RAs)}}, each representing a distinct resource provider with access to and local knowledge of its own set of resources---referred to as \textbf{\textit{Capacity}} in this paper. RAs can dynamically join or leave the network and are primarily responsible for exposing their resources and collaboratively working with other RAs to discover and allocate suitable resources from their capacities for incoming applications. 

Swarmchestrate supports microservices-based applications, each comprising $m$ components with specific resource requirements. Upon submission, the application request is broadcast to all RAs, who then initiate a collaborative process to identify and reach consensus on the most suitable resources. This paper focuses specifically on this aspect. For a detailed account of Swarmchestrate’s architecture and the integration of this work, see~\cite{UllahMAKKDMWK25}; for its broader vision, refer to~\cite{kiss2024swarmchestrate}. 

\subsection{Distributed task allocation}
Distributed task allocation---a well-studied problem in Multi-Agent Systems---is characterised along three dimensions: (1) single- vs. multi-agent tasks---whether a task requires one or multiple agents to execute; (2) single- vs. multi-task agents---whether an agent can perform one or multiple tasks; and (3) instantaneous vs. time-extended---whether tasks are allocated immediately or over a period of time.

In our context, the problem is framed as an independent, single-agent, multi-task, instantaneous task allocation. Specifically, the resource requirements of each microservice (i.e., a task) are independent, with no interdependencies, and each is allocated to at most one RA representing a capacity. Lastly, allocation decisions are made on the fly, as application requests arrive. To solve such problems, two classes of methods are commonly employed: Distributed Constraint Optimisation Problems (DCOP) and Market-Based Auctions~\cite{Ahmadoun22}.

DCOPs are a distributed form of constraint optimisation in which multiple agents, each with limited knowledge and control, collaboratively compute assignments that optimise a global objective while satisfying local constraints~\cite{FiorettoPY18}. DCOPs have been applied to task allocation and other domains such as scheduling, supply chains, and smart grids~\cite{FiorettoPY18}. Once the problem is formulated as a DCOP, several algorithms, such as complete or approximate algorithms, can be utilised~\cite{FiorettoPY18}. 

Market-based auction methods, on the other hand, follow a three-stage process: task announcement to inform agents, bidding by agents based on task-specific utility evaluations, and winner selection, where the task is assigned to the agent with the highest bid~\cite{Mosteo10}. A well-known example of such methods is the Consensus-Based Bundle Algorithm (CBBA)~\cite{ChoiBH09}. This algorithm was first proposed to solve robot task allocation; however, it has also been widely used in other domains (e.g., autonomous vehicles~\cite{luo2025demand}, and satellite planning~\cite{aguilar2025decentralized}) due to its effectiveness in handling multi-agent, multitask scenarios.


In this article, we choose to use CBBA over DCOP for three key reasons~\cite{KumarFP09}. First, CBBA offers a simpler and more intuitive problem formulation compared to the complexity of modelling required for DCOP. Second, CBBA generally incurs lower communication overhead and shorter execution times. Third, CBBA demonstrates better scalability when handling a larger number of agents. More details on the adaptation of CBBA can be found in Section~\ref{sec:cbba}.

\subsection{Related work}
The landscape of resource selection, allocation, and orchestration is diverse. Since this paper focuses on distributed resource selection within the broader context of orchestration, we limit our discussion to such related work. For a detailed review, refer to relevant survey papers, e.g.~\cite{patsias2023task,ullah2023orchestration}.

Hierarchical approaches have been proposed in various studies, such as mF2C~\cite{masip2021managing}, Oakestra~\cite{bartolomeo2023oakestra}, where the overall infrastructure has been distributed into independent logical layers of the Cloud-Edge continuum. Each of these layers is controlled by an independent orchestrator, where communication and coordination are restricted among these orchestrators for resource selection and deployment decisions.  

P2P models, such as HYDRA~\cite{jimenez2020hydra} and Caravela~\cite{pires2021distributed}, implement a P2P overlay network in which each computational node acts as both a resource and an orchestrator, enabling distributed resource discovery and microservice execution. The resource discovery in these mechanisms relies on standard P2P protocols, such as distributed hash tables and lookup algorithms; however, these approaches lack a utility-based optimisation mechanism for resource selection.


Several approaches leverage AI-based techniques~\cite{schneider2021distributed, li2025adaptive, ullah2023optimizing}. For instance, Schneider et al.~\cite{schneider2021distributed} proposed a distributed self-learning-based coordination approach using DRL. However, it relies on a centralised training process. Li et al.~\cite{ li2025adaptive} introduced a hybrid approach combining Graph Neural Networks (GNN) with collaborative multi-agent reinforcement learning (MARL), where local agents make decisions at neighbourhood-level, and a global orchestrator manages system-wide coordination. Similarly, Ullah et al.~\cite{ullah2023optimizing} adopted a hybrid approach using Markov Decision Processes (MDP) and Double Deep Q-Networks (DDQN) to optimise computation offloading and resource allocation in edge computing.

Other studies~\cite{markus2025proximity, ding2021multi, mudassar2020decentralized} focus on grouping resources or agents based on specific characteristics such as proximity or task dependencies. These approaches used different techniques, e.g. clustering algorithms in~\cite{markus2025proximity}, MARL in~\cite{ding2021multi}, and manual group formation in~\cite{mudassar2020decentralized}, where a designated agent coordinates the group based on participating nodes. However, in all cases, each group is collectively responsible for executing all tasks of an application.

In the same realm, Zeinab et al.~\cite{nezami2021decentralized} proposed a more cooperative mechanism in which each agent follows a two-step process: (1) propose a resource allocation plan for a given request and share it with all neighbours, and (2) select one plan from all received proposals, leading to a final deployment plan consist of aggregating all selected proposals. This approach, however, assumes full knowledge of all neighbour nodes and their capacities. Finally, gossip-based methods have also been explored---for instance, Sim et al.~\cite{sim2022cooperative} adapted the KM gossip algorithm---for discovering fog nodes and allocating resources in a fully distributed manner.

In contrast to these methods, our CBBA-based approach eliminates the need for any centralised training, predefined grouping, neighbour or global knowledge, instead enabling a distributed, utility-driven resource allocation that can adapt dynamically to the heterogeneous and large-scale nature of the cloud–edge continuum.

\section{Methods} \label{sec:methods}
This section presents the three core technical methods that underpin our system's approach to resource selection. The first two, described in Section~\ref{sec:centralised}, follow a centralised architecture and serve as baselines for comparison. Section~\ref{sec:cbba} further introduces our novel adaptation of CBBA as a distributed resource selection mechanism.

\subsection{Centralised Resource Selection}\label{sec:centralised}
In our earlier work~\cite{UllahMAKKDMWK25}, the overall architecture of the Swarmchestrate orchestration framework follows a distributed architecture; however, the resource selection part remained fully centralised. This process began with a single RA receiving the application submission request, thereby assuming the role of a central controller responsible for the following steps:
\begin{enumerate}
    \item Broadcasting resource requirements of the application to other RAs.
    \item Collecting bids (resource offers) from RAs. Each offer indicates that an RA can fulfil the resource needs of a specific microservice (component/task) of the submitted application from its available capacity.
    \item Generating valid allocations by combining received offers. A valid allocation satisfies all resource requirements of the application, meaning all its microservices are mapped to compatible RAs' capacities.
    \item Evaluating each valid allocation using a cost function, rank them, and select the best allocation.
\end{enumerate}

The cost function, developed in~\cite{UllahMAKKDMWK25}, computes the weighted sum of several QoS attributes (in normalised form). It also incorporates weighting factors that allow prioritisation of specific QoS attributes over others. However, in this paper, we use a balanced configuration---assigning equal weights to all QoS factors. Regarding compiling valid allocations from resource offers, we consider the following two variants:
\begin{enumerate}
    \item Centralised (All enumerations): This variant, as in~\cite{UllahMAKKDMWK25}, enumerates resource offers to create all possible valid allocations. Once all possible allocations are obtained, they are scored using the cost function, and the allocation with the minimum cost is used to run the application. This method has the benefit of obtaining optimal allocations. Nevertheless, it can be time-consuming, especially if the number of valid allocations is large.
    \item First-Fit: This variant is based on the use of the first-fit heuristic method to build valid allocations incrementally by selecting the first offers that satisfy remaining requirements. For instance, if an initial offer covers one microservice, the heuristic constructs a valid allocation using the first matching offers it encounters for the other microservices. This method is significantly faster; however, it may miss some valid allocations, as it does not exhaustively explore all possibilities.
\end{enumerate}
Throughout the remainder paper, we refer to these two approaches as \textbf{\textit{Centralised}} and \textbf{\textit{First-Fit}}, respectively.

\subsection{CBBA-based Distributed Resource Selection}
\label{sec:cbba}
The goal of the distributed resource selection mechanism is to eliminate the reliance on a central controller and to address the inherent limitations of centralised approaches, such as latency, scalability bottlenecks, and single points of failure. In the distributed model, RAs collaborate without central coordination to select resources for incoming applications. Each RA independently assesses its local capacity and engages in lightweight communication with peers to collectively determine a suitable allocation. To enable such collaboration among RAs, we utilised the CBBA~\cite{ChoiBH09}---a well-established auction-based mechanism designed for distributed task allocation. 

CBBA mainly combines elements of consensus and greedy selection to ensure conflict-free allocations in a distributed setting. CBBA offers a structured approach, where agents iteratively build bundles of preferred tasks based on the evaluations of local utility and then use a consensus phase to resolve conflicts and reach agreement across the network. This makes it particularly well-suited for the cloud-edge continuum, where resource agents operate with partial knowledge and must coordinate efficiently without a central control.

In our context, a set of resource agents $\mathcal{RA}$ bid on the resources required to host microservices $\mathcal{MS}$, treating each microservice as an individual task. Each RA makes bidding decisions based on the local information it holds, namely, its utility for hosting a given microservice and its current view of which agent is winning which task. Through this bidding and consensus process, agents collaboratively and efficiently coordinate resource selection without a central controller. 

Overall, the CBBA algorithm operates in two iterative phases---bidding and consensus---which repeat until convergence is reached. In our context, convergence occurs when all RAs agree on a consistent and conflict-free allocation of resources from their local capacities, meaning that each microservice is assigned to exactly one RA, and no further updates or reassignments take place across the network in subsequent rounds. The following sections provide a detailed explanation of the relevant aspects:

\paragraph{Bidding phase}

Each agent computes its local utility $\mathcal{U} = u_{im},\ i \in [0, |\mathcal{RA}|],\ m \in [0, |\mathcal{MS}|]$, where each entry represents the agent’s utility for a specific microservice. Using this, agents construct an ordered sequence---bundle---of tasks, prioritising those that yield the highest utility. This bundle guides their bidding decisions. The utility function here is the same as the cost function discussed earlier in Section~\ref{sec:centralised}, with one difference, i.e., it is inverted, since CBBA operates under a maximisation framework rather than minimisation.

\paragraph{Consensus phase}
Each RA exchanges messages with its peers to update their bundle based on newly obtained information. A message sent by an agent $a_i \in \mathcal{RA}$ mainly includes the following three elements~\cite{ChoiBH09}:

\begin{enumerate}
    \item $y_i \in \mathbb{R}_+^{|\mathcal{MS}|}$, the utility values of the current winning bids for all microservices;
    \item $z_i \in \mathcal{RA}^{|\mathcal{MS}|}$, the identity of the current winning agent for each task (i.e., the RA holding the highest bid); 
    \item $s_i \in \mathbb{R}^{|\mathcal{RS}|}$, the timestamp indicating when agent $a_i$ last obtained information about the winning bid for a task. This timestamp enables agents to resolve bid conflicts by prioritising the most recent updates.
\end{enumerate}

When two agents $a_i$ and $a_j$ exchange a message, $a_i$ updates its information with $a_j$'s information on a microservice $m$ if:
\begin{enumerate}
    \item $s_j > s_i$, i.e., the information from agent $a_j$ is latest; 
    \item $s_j = s_i$ and $u_{im} < u_{jm}$, i.e., both agents possess the most up-to-date information, and agent $a_j$’s utility for microservice $m$ is higher than that of agent $a_i$.
    \item $s_j = s_i$, $u_{im} = u_{jm}$, and $i > j$, i.e., when both agents have up-to-date information and identical utility values, the tie is resolved by using their agent IDs.
\end{enumerate}

\paragraph{Convergence properties}
Convergence is achieved when all agents share an identical winners list, i.e., $z_i = z_j, \forall i,j \in \mathcal{RA}$. This state indicates that every agent has the same view of task assignments, and no further changes will occur. Convergence is guaranteed if the utility function satisfies the \emph{diminishing marginal gains} property~\cite{ChoiBH09}, meaning that the score assigned to a task remains unchanged when additional tasks are added to the bundle. Furthermore, CBBA provides a theoretical performance guarantee, achieving at least 50\% of the optimal solution in the worst-case scenario~\cite{ChoiBH09}.


\paragraph{Key modifications}
The CBBA algorithm was originally developed for task allocation in the robotics domain. To adapt it to the Cloud–Edge resource selection problem, several modifications were made. First, the original formulation assumes mobile robots, where task allocation depends on their geographic location. In our context, computational resources across the Cloud–Edge continuum are static and do not involve mobility. Consequently, we removed the mobility aspect from the algorithm. Instead, location is treated as an optional constraint attached to certain microservices, restricting resource selection only to those that satisfy this requirement.

Similarly, the original CBBA imposes a geographic constraint on communication---robots can only exchange messages if they are within proximity. Since no such constraint exists in our setting, we simplified the model, where all RAs can communicate directly with one another. More complex communication topologies can be considered in future work.

Another modification concerns the tie-breaking mechanism. As in the original algorithm, ties are resolved first by timestamp (preferring the most up-to-date information), then by bid value (preferring higher bids), and finally by agent ID as a last resort. However, before applying the agent ID rule, we introduce an additional tie-breaking step based on price discount, favouring capacities that offer better cost reductions when multiple microservices are allocated to them.
\section{Experimental Setup and Results}
\label{sec:results}
This section provides a comprehensive evaluation of the proposed method against the two baseline methods. We first outline the experimental setup, detailing key aspects such as resource capacities, applications, and the QoS attributes used to evaluate the quality of resource allocations. Using this setup, we then evaluate all three approaches, focusing specifically on their efficiency, performance, and scalability.

\subsection{Applications, Capacities and QoS} \label{sec:generation}
The core elements of our experiments are applications and capacities. 
An application comprises $m$ microservices, each with \#CPU and \#RAM requirements, an execution time bound, and optionally a location constraint, specifying where computation should occur. A capacity represents a set of resources provided by a cloud or edge provider and accessed via an RA. Each capacity is characterised by its quota of \#CPU, and \#RAM, a physical location, and a type (cloud or edge). Cloud capacities can serve multiple applications, while Edge capacities handle only one at a time.

Each capacity has four associated QoS attributes: bandwidth, latency, price, and energy consumption. Price and energy consumption are calculated on a per unit resource per unit time; for example, the cost to run a microservice is computed as $price = p \times (\#CPUS + \#RAM) \times \text{Running time}$, where p is the unit price of the capacity, and \#CPU and \#RAM are the microservice’s resource demands. The same formula applies to energy consumption. Capacities additionally feature a discount factor applied when multiple microservices from the same application are allocated to that capacity, meaning it incentivises consolidation. These QoS attributes inform the cost function in centralised methods and the utility function in the CBBA-based method to evaluate valid allocations.

For all experiments in this paper, similar to our prior work~\cite{UllahMAKKDMWK25} and related studies~\cite{
jimenez2020hydra, pires2021distributed}, application and capacity specifications (Tables~\ref{tab:app} and~\ref{tab:capacity}) were generated randomly. Each experiment is composed of the following: the resources to run the microservices of $a$ applications are selected from the available $c$ capacities one after the other. Each experiment is repeated five times for applications and capacities. Once a microservice is allocated to a capacity, the capacity available quota is reduced accordingly by the microservice resource requirements (\#CPU and \#RAM).

\begin{table}[]
\caption{Application specifications.}

\label{tab:app}
\centering
\begin{tabular}{l|lll}
Parameter & Value &  &  \\ \cline{1-2}
Microservices & min:1, max: 5 &  &  \\
CPU & min: 1, max: 6 &  &  \\
RAM (GB) & min: 1, max: 6 &  &  \\
Storage (GB) & min: 1, max: 10 &  &  \\
Location & EU (20\%), US (20\%), Asia (20\%), Worldwide (40\%) &  &  \\
\begin{tabular}[c]{@{}l@{}}Running Time \\ (unit of time)\end{tabular} & min:1, max: 30 &  &  \\ \cline{1-2}
\end{tabular}
\end{table}

\begin{table}[]
\caption{Capacities specifications.}
\label{tab:capacity}
\centering
\begin{tabular}{l|l|ll}
Parameter                                                    & Value                                                                                                              &  &  \\ \cline{1-2}
Type                                                         & Cloud or Edge                                                                                                      &  &  \\
\begin{tabular}[c]{@{}l@{}}CPU\\ $2^n$\end{tabular}          & \begin{tabular}[c]{@{}l@{}}Cloud: min: 4, max: 10\\ Edge: min: 1, max: 5\end{tabular}                              &  &  \\
\begin{tabular}[c]{@{}l@{}}RAM\\ $2^n$ (GB)\end{tabular}     & \begin{tabular}[c]{@{}l@{}}Cloud: min: 4, max: 10\\ Edge: min: 1, max: 5\end{tabular}                              &  &  \\
\begin{tabular}[c]{@{}l@{}}Storage\\ $2^n$ (GB)\end{tabular} & \begin{tabular}[c]{@{}l@{}}Cloud: min: 2, max: 13\\ Edge: min: 2, max: 10\end{tabular}                             &  &  \\
Location                                                     & EU, US, Asia                                                                                                       &  &  \\
Price (\$)                                                   & \begin{tabular}[c]{@{}l@{}}US: min: 0.15, max: 1\\ EU: min: 0.1, max: 0.8\\ Asia: min: 0.05, max: 0.7\end{tabular} &  &  \\
Energy (W)                                                   & min: 1, max: 10                                                                                                    &  &  \\
Bandwidth (Mbps)                                                 & min: 100, max: 1000                                                                                                &  &  \\
Latency (ms)                                                 & min: 50, max: 200                                                                                                  &  &  \\
Discount                                                     & min: 0, max: 1                                                                                                     &  &  \\ \cline{1-2}
\end{tabular}
\end{table}

\subsection{Computation time}
\label{sec:time}
\begin{figure}
    \centering
    \includegraphics[width=\linewidth]{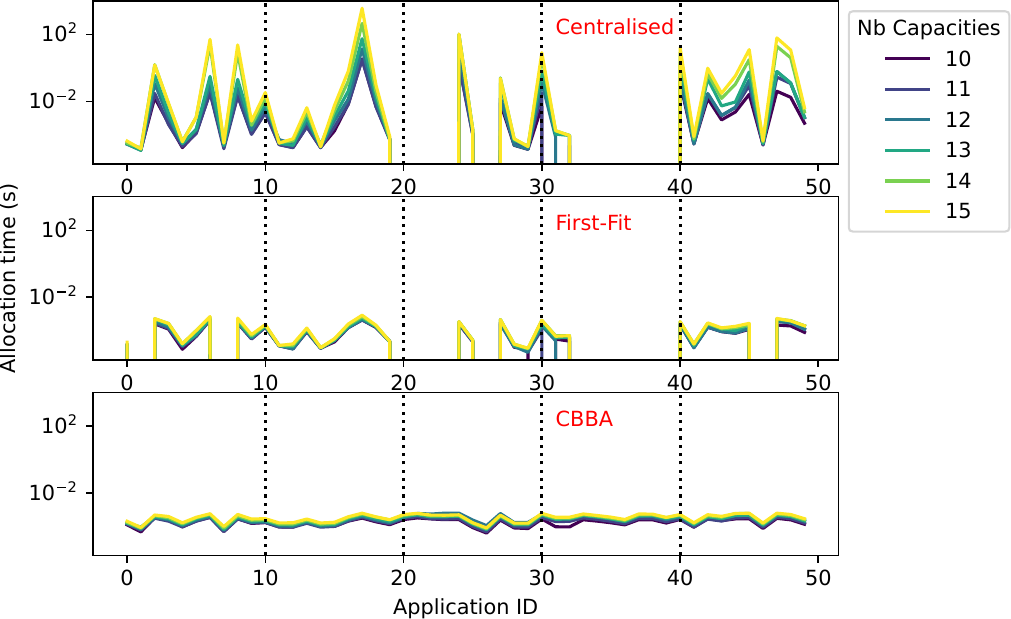}
    \caption{Allocation time (in log-scale) by application with varying number of capacities (ranging from 10 to 15). Black dotted lines represent the five repetitions.}
    \label{fig:time}
\end{figure}

We conducted experiments using a fixed set of $10$ applications and a varying number of capacities, ranging from $10$ to $15$, with each configuration repeated over five independent runs. The computation times for all three methods are summarised in Figure~\ref{fig:time}. Across all methods, increasing the number of available capacities from $10$ to $15$ resulted in longer computation times. The centralised method, however, is most sensitive to such an increase. Specifically, its average computation time rises from $0.15s$, with $10$ capacities, to $81s$ with $15$ capacities. This difference is even more prominent in the worst-case scenario, where the time grows from $3.5s$ to over $3,800s$ when capacities are increased from $10$ to $15$.

In contrast, both First-Fit and CBBA maintain consistently low computation times with an average around $10^{-4}s$. For the First-Fit, this efficiency stems from the smaller number of valid allocations explored compared to the exhaustive enumeration in the centralised approach. For CBBA, the computation time primarily depends on the number of messages exchanged among RAs, which scales with the number of capacities; however, it remains minimal compared to the other methods.

\subsection{Failed cases} \label{sec:failed}

For both centralised and First-Fit approaches, the allocation process begins with a pre-selection step that compiles potential allocations based on the offers from RAs. This step occurs before the evaluation of allocations. Consequently, if no valid allocation can be formed during this stage, the evaluation step is skipped, and no computation time is recorded, hence the gaps in the results (Figure~\ref{fig:time}), representing the failed allocations. In contrast, CBBA does not employ the pre-processing step because it does not rely on offers. Instead, it operates through iterative bidding and consensus phases, regardless of whether a final valid allocation is found or not. As a result, no gaps in the CBBA’s reported computation times. 

\begin{figure}
     \centering
     \includegraphics[width=\linewidth]{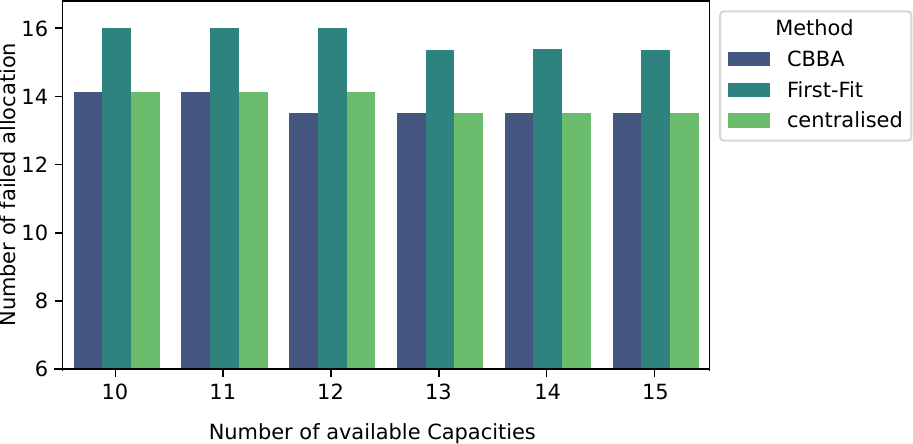}
     \caption{Number of failed allocations for $50$ applications under varying levels of capacities.}
     \label{fig:failure}
 \end{figure}

Figure~\ref{fig:failure} presents the number of failed cases for each method across the full evaluation set of $50$ applications (i.e., $10$ applications repeated $5$ times), with the number of available capacities varying from $10$ to $15$. Among the three methods, First-Fit exhibited the highest failure rate. For instance, when the number of available capacities was below $13$, approximately $32\%$ of applications could not be allocated. This rate dropped slightly to $30\%$ when the number of capacities increased to between $13$ and $15$. In contrast, the centralised method consistently showed a low failure rate in all cases, with one notable exception. This consistency is largely due to its exhaustive enumeration of all valid combinations; thus, if a failure occurs, it is only because no allocation is truly possible given the available resources. This behaviour makes the centralised method generally more reliable in avoiding failures in allocation.

Interestingly, CBBA matches or even outperforms the centralised method in terms of failure rate. In most scenarios, the number of failures is identical for both methods. However, in the specific case of $50$ applications with $12$ capacities, CBBA records fewer failures. This difference arises from a single sub-optimal allocation made by CBBA for one application (see Section~\ref{sec:allocation}). Consequently, this sub-optimal allocation preserved enough capacity to enable a subsequent application---one that failed under the centralised method---to be successfully allocated under CBBA. Thus, a minor inefficiency in CBBA’s process inadvertently improved the overall allocation outcome.

\subsection{Allocation cost}
\label{sec:allocation}
The cost function, as discussed in Section~\ref{sec:methods}, is used to measure the quality of an allocation based on four QoS. In this section, we analyse the qualities of these allocations obtained using all methods based on their cost value.

\begin{figure}
    \centering
    \includegraphics[width=\linewidth]{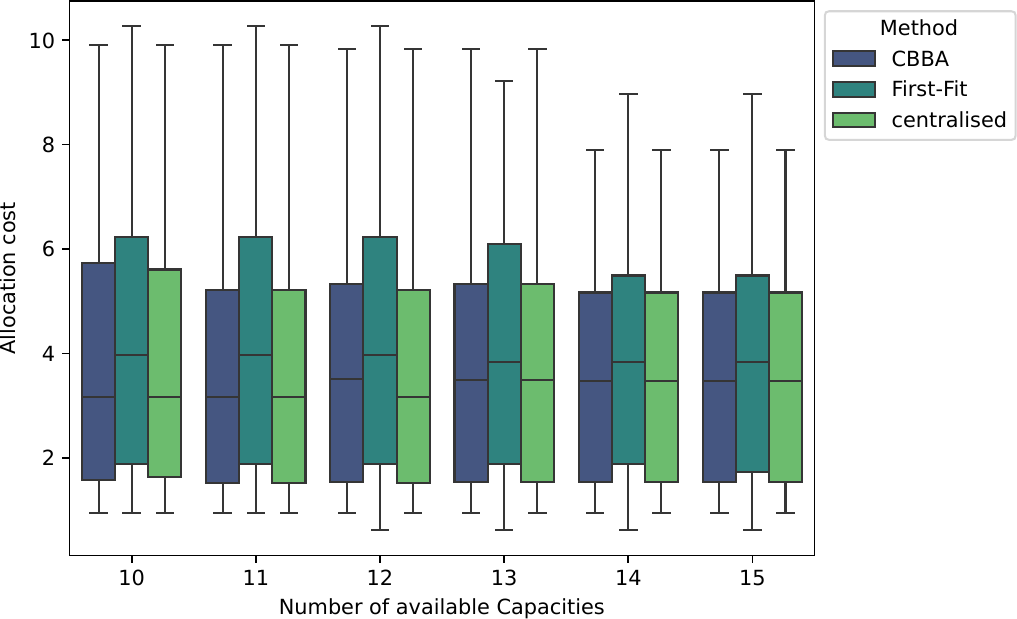}
    \caption{Allocation cost for $50$ applications.}
    \label{fig:cost}
\end{figure}

Figure~\ref{fig:cost} presents the cost distributions for all three methods, evaluated on the same settings as discussed in Section~\ref{sec:time} (i.e., $5$ times $10$ applications with capacities ranging from $10$ to $15$). Overall, the cost differences among the three methods are minimal. However, First-Fit displays a noticeably wider spread in costs compared to the other two approaches, indicating greater variability in its selection outcomes. To assess statistical significance, we conducted Kolmogorov–Smirnov two-sample tests comparing (i) centralised vs. CBBA cost samples and (ii) centralised vs. First-Fit cost samples. In both cases, the results were not statistically significant, meaning we cannot reject the null hypothesis that the samples come from the same distribution. Hence, overall, the three methods exhibit similar performances in terms of the overall cost for each application.

\begin{figure}
    \centering
    \includegraphics[width=\linewidth]{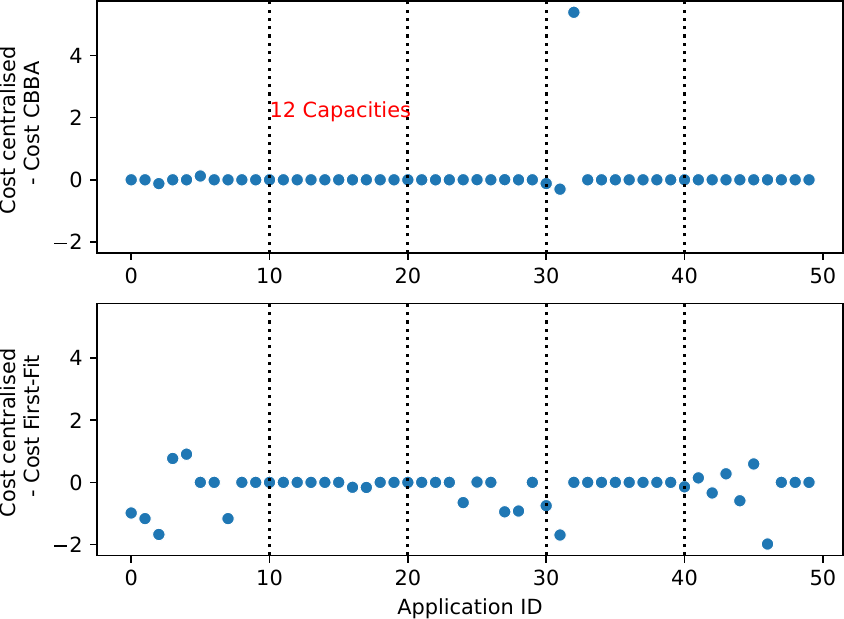}
    \caption{Difference of allocation costs for individual applications with $12$ available capacities.}
    \label{fig:cost_indiv}
\end{figure}

Figure~\ref{fig:cost_indiv} provides a closer look at the per-application cost differences when $12$ capacities are available (Results for other settings are in the supplementary materials~\cite{dataCBBA}). The black dotted lines indicate the five repetitions of $10$ applications, meaning that the capacities are reset after each set of $10$.

The top plot compares the cost of centralised and CBBA allocations. Out of the $50$ applications, only $5$ show a different cost, indicating a different allocation. Among these, CBBA yields a higher cost than the centralised variant in $3$ cases. This is because CBBA is not an exact algorithm and thus may not find the optimal allocation. However, interestingly, in $2$ cases, CBBA outperforms the centralised method, despite the latter exhaustively enumerating all allocations. This is possible due to that CBBA produced non-optimal allocations for earlier applications (e.g., $\#30$ and $\#31$), which inadvertently preserved resources that enabled a valid allocation for a later application (e.g., $\#32$ in this case) where the centralised method failed.

The bottom plot shows the difference in cost between the centralised and First-Fit methods. Here, the number of differences is noticeably larger than in the CBBA comparison. This is because First-Fit’s allocation process may miss valid allocations that would yield a lower cost, resulting in more frequent sub-optimal outcomes. As with CBBA, sub-optimal allocations result in a difference of resources available and explain the variability observed in the results of Figure~\ref{fig:cost_indiv}.

\subsection{QoS analysis}
\label{sec:qos_analysis}
\begin{figure}
    \centering
    \includegraphics[width=\linewidth]{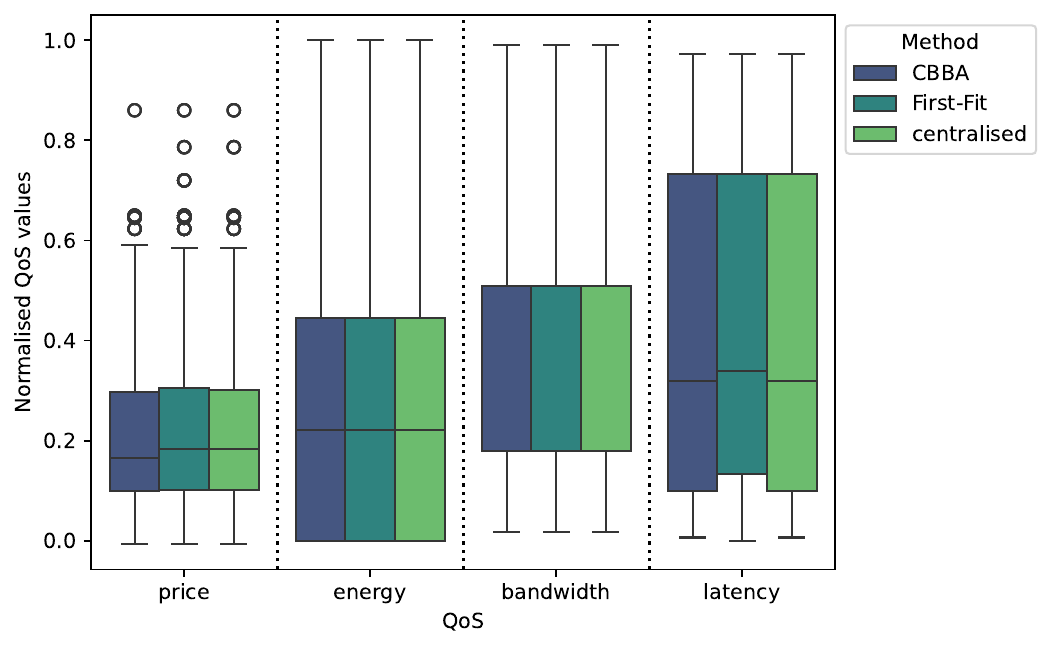}
    \caption{QoS values across $50$ applications for all capacities ($10$ to $15$). Lower value means better QoS.}
    \label{fig:qos}
\end{figure}

It is evident from Section~\ref{sec:allocation}---particularly in Figure~\ref{fig:cost}---that the overall allocation cost for all three methods is broadly similar. This cost is computed as the sum of four QoS attributes: price, energy consumption, bandwidth, and latency. While the total cost is comparable across methods, the individual QoS values may still differ. To examine these differences, we further analyse each QoS attribute separately. The results are presented (in normalised form) in Figure~\ref{fig:qos}. The lower value indicates better QoS. 

Consistent with Figure~\ref{fig:cost}, the results reveal minimal differences in the distributions, suggesting that the observed similarity in total cost is also reflected in each QoS dimension. Furthermore, similar to our earlier analysis of overall cost distributions using Kolmogorov–Smirnov two-sample tests, we applied the same tests to each QoS attribute, comparing centralised vs. CBBA and centralised vs. First-Fit. 

For centralised vs. CBBA, none of the tests were statistically significant, indicating that we cannot reject the null hypothesis that both sets of samples are drawn from the same distribution. This result confirms that CBBA delivers QoS values comparable to those of the centralised method. Similarly, for centralised vs. First-Fit, three out of the four QoS attribute tests---price, energy, and latency---were not statistically significant, indicating comparable performance between the two methods for these metrics. However, the bandwidth test yielded a significant result, allowing us to reject the null hypothesis that the two sets of values are drawn from the same distribution. Since the centralised allocation consistently produced higher bandwidth values than First-Fit, it outperforms First-Fit in this specific QoS dimension.

\subsection{Large scale experiments}
The experiments presented in the previous sections are relatively small-scale, with a few number of applications and capacities. It is evident from the results that the CBBA performs on par with the centralised variant, which exhaustively enumerates all valid allocations. However, such exhaustive enumeration quickly becomes impractical due to its high computational cost, making it unsuitable for larger problem sizes. For instance, even with as few as $15$ capacities, computing the optimal allocation can take up to an hour.

\begin{figure}
    \centering
    \includegraphics[width=\linewidth]{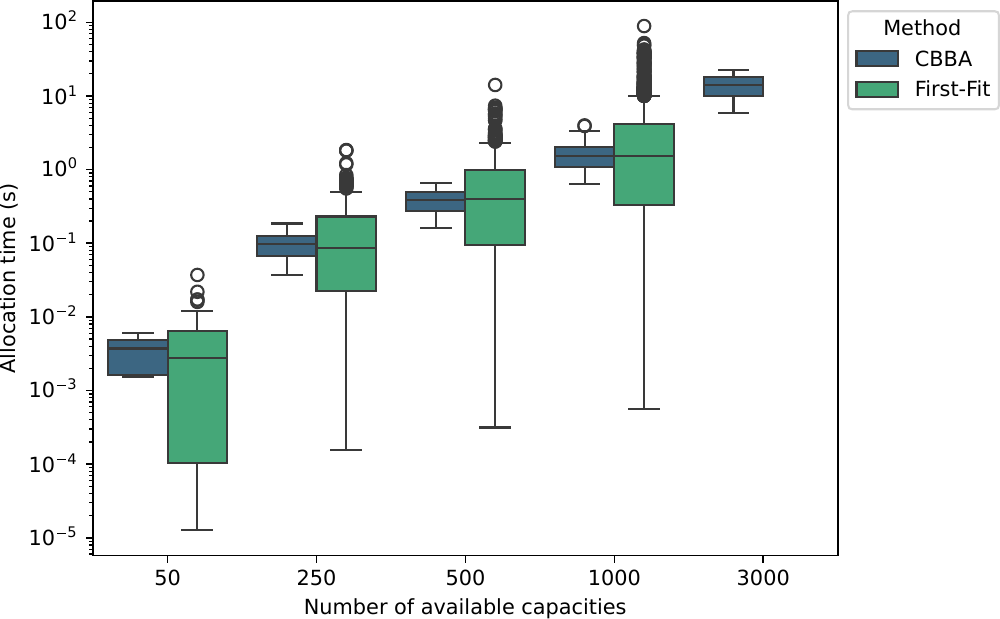}
    \caption{Allocation time for up to $3{,}000$ capacities): First-Fit vs CBBA. First-Fit timed out for $3{,}000$ capacities}
    \label{fig:time_scale}
\end{figure}

\begin{figure}
    \centering
    \includegraphics[width=\linewidth]{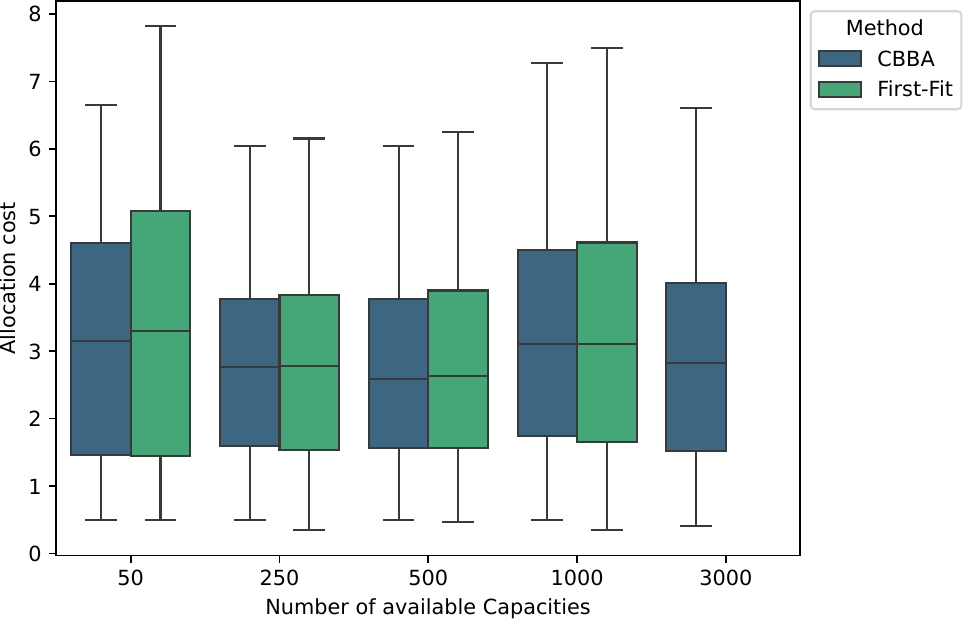}
    \caption{Allocation cost: for up to $3{,}000$ capacities): First-Fit vs CBBA. First-Fit timed out for $3{,}000$ capacities.}
    \label{fig:cost_scale}
\end{figure}

To evaluate the scalability perspective, this section focuses on the two practical methods---First-Fit and CBBA. For this purpose, we designed five scenarios, each repeated five times, varying the number of applications and capacities as follows: (1) $10$ applications, $50$ capacities, (2) $50$ applications, $250$ capacities, (3) $100$ applications, $500$ capacities, (4) $500$ applications, $1{,}000$ capacities, (5) $1{,}000$ applications, $3{,}000$ capacities. The results for computation time (log scale) and allocation cost are presented in Figures~\ref{fig:time_scale} and \ref{fig:cost_scale}, respectively.

Both methods achieve comparable median times (Figure~\ref{fig:time_scale}) across most scenarios. However, in the largest scenario, First-Fit failed to complete within a reasonable time, and only CBBA results are reported. Notably, First-Fit exhibits substantially higher variability. For instance, with $1{,}000$ capacities, First-Fit’s maximum computation time reached $89.2$ seconds, compared to just $3.98$ seconds for CBBA. This variability stems from the way First-Fit generates allocations---by enumerating allocations from combinations of resource offers---making its runtime highly sensitive to both the number of microservices per application and the number of available capacities. In contrast, CBBA’s runtime depends primarily on the number of capacities, which affects the number of agents and the volume of inter-agent communications; however, it is less influenced by the number of microservices. Consequently, CBBA scales more effectively; even with $3{,}000$ capacities, it achieves a median computation time of only $13.9$ seconds.

In terms of allocation cost (Figure~\ref{fig:cost_scale}), CBBA and First-Fit demonstrate comparable performance across all scenarios. To verify this, we conducted Kolmogorov–Smirnov two-sample tests on the cost samples from both methods for each setting. None of the tests were statistically significant, indicating that we cannot reject the null hypothesis that the samples are drawn from the same distribution. Thus, both methods achieve similar cost performance at scale. Notably, unlike in the smaller-scale experiments presented in Section~\ref{sec:allocation}, no allocation failures were observed for any application in these large-scale settings.

\section{Conclusion} \label{sec:conclusion}
Our adaptation of CBBA to the cloud–edge continuum enables scalable, decentralised resource selection, matching the optimality of exhaustive search and the efficiency of heuristic methods—without their inherent limitations. Unlike existing approaches, our method requires no training, predefined grouping, or global/neighbour knowledge, enabling fully decentralised decision-making that scales to large and heterogeneous cloud-edge infrastructures. Through extensive evaluation, we demonstrated that our approach delivers allocations with performance comparable to centralised heuristics in terms of cost and QoS, while significantly reducing computation time---achieving timely results even in large-scale scenarios where exhaustive search is intractable and centralised methods are up to 30 times slower. These results highlight CBBA’s suitability for real-world, large-scale cloud–edge orchestration systems where responsiveness and scalability are critical. 

\section*{Acknowledgements}
This work is co-funded by the European Union’s Horizon Europe programme under grant agreement No. 101135012, and by UK Research and Innovation (UKRI) under grant agreement No. 10102651, as part of the project Swarmchestrate: Application-level Swarm-based Orchestration Across the Cloud-to-Edge Continuum.
\bibliographystyle{IEEEtran}
\bibliography{IEEEabrv,references_sw}

\end{document}